\documentstyle[prd,eqsecnum,preprint,tighten,aps]{revtex}
\begin{document}
\draft
  \title{Probing the Brans-Dicke Gravitational Field by Cerenkov Radiation}
\author{G. Lambiase$^{a,b}$\thanks{E-mail: lambiase@sa.infn.it}}
\address{$^a$Dipartimento di Fisica "E.R. Caianiello"
 Universit\'a di Salerno, 84081 Baronissi (Sa), Italy.}
 \address{$^b$Istituto Nazionale di Fisica Nucleare, Sez. di Napoli, Italy.}
\date{\empty}
\maketitle
\begin{abstract}

The possibility that a charged particle propagating in a
gravitational field described by Brans-Dicke theory of gravity
could emit Cerenkov radiation is explored. This process is
kinematically allowed depending on parameters occurring in the
theory. The Cerenkov effect disappears as the BD parameter
$\omega\to \infty$, i.e. in the limit in which the Einstein
theory is recovered, giving a signature to probe the validity of
the Brans-Dicke theory.

\end{abstract}

\section{Introduction}
\setcounter{equation}{0}

Alternative theories of gravity \cite{will} represent an extension
of General Relativity in which the gravitational interaction is
mediated by a tensor field and one or several scalar fields, in
general massless. The Brans-Dicke (BD) theory of gravity
\cite{brans} is, may be, the most famous prototype of such
theories which incorparates Mach's principle and Dirac's large
number hypothesis (see, for example, \cite{weinberg}) by means of
a nonminimal coupling between the geometry and a scalar field
$\phi$, the BD scalar. The scalar field rules dynamics together
with geometry and induces a variation of the gravitational {\it
coupling} with time and space through the relation
$G_{eff}=1/\phi$. The Newtonian gravitational constant $G_N$ is
recovered in the limit $\phi\to constant$. As recent experiments
\cite{pioneer,nandi-rev} have shown, a variation of the Newton
constant on astrophysical and cosmological sizes and time scale
seems to have been confirmed.

During last years there has been a growing interest for the
Brans-Dicke gravity, as well as for the alternative theories of
gravity \cite{will,damour}, due to the following reasons: $i)$ The
presence of scalar fields seems to be unavoidable in supersting
theory \cite{green}. $ii)$ BD gravity follows from Kaluza-Klein
theory in which the compactified extra dimensions is essential for
generating scalar fields \cite{cho}. $iii)$ BD theory is an
important ingredient in the scenario of extended and hyperextended
inflation \cite{la,kolb,steinardt,liddle}. Besides, constraints of
the coupling constant of BD theory are derived in relation to the
observations of gravitational waves from compact binaries
\cite{will1}, whereas investigations concerning the collapse to
black holes in the framework of BD has been carried out in
\cite{scheel}. Recently, consequences of the BD gravitational
field on neutrino oscillations and Sagnac effect have been
analyzed in \cite{gaetano} and \cite{nandi}, respectively. In this
paper we suggest a new physical effect which involves the Cerenkov
emission by charged particles propagating in curved space-time.

\section{The BD Theory of Gravity}
The effective action describing the interaction of the scalar
field $\phi$ nonminimally coupled with the geometry and the
ordinary matter is given by \cite{brans}
\begin{equation}\label{action}
{\cal A}=\int d^4x \sqrt{-g}\left[\phi R-\omega
\frac{\partial_{\mu}\phi\partial^{\mu}\phi}{\phi}+\frac{16\pi}{c^4}
{\cal L}_{m}\right]\,,
\end{equation}
where $R$ is the scalar curvature, ${\cal L}_{m}$ is the matter
contribution in the total Lagrangian density. The constant
$\omega$ can be positive or negative. In what follows we will
assume $\omega>0$. On an experimental setting, $\omega$ is
determined by observations and its value can be constrained by
classical tests of General Relativity: the light deflection, the
relativistic perihelion rotation of Mercury, and the time delay
experiment, resulting in reasonable agreement with all available
observations thus far provided $\omega \geq 500$
\cite{will,will2}. The most recent bounds on $\omega$ comes from
Very-Long Baseline Radio Interferometry (VLBI) experiment and it
is given by $\omega>3000$ \cite{will-rev}. On the other hand,
bounds on the anisotropy of the cosmic microwave background
radiation (CMB) give the upper limit $\omega\leq 30$ \cite{la}.
Nevertheless, it must be mentioned that results obtained in Refs.
\cite{xue}, in which the BD theory during the radiation matter
equality has been investigated, seem to indicate that at the
moment there is no observable effect appearing from CMB. It is
important to point out that Einstein's theory is recovered as
$\omega\to\infty$. In this limit, the BD theory becomes
indistinguishable from General Relativity in all its predictions
(this is not always true, see for example
\cite{barrows,faraoni,sen}).

Variation with respect to the metric tensor $g_{\mu\nu}$ and the
scalar field $\phi$ yield the field equations \cite{brans}
\begin{equation}\label{2.1}
  R_{\mu\nu}-\frac{1}{2}\, g_{\mu\nu}R=
  \frac{\omega}{\phi^2}\left(\phi_{,\mu}\phi_{,\nu}-\frac{1}{2} g_{\mu\nu}
  \phi_{,\alpha}\phi^{,\alpha}\right)
  +\frac{1}{\phi}\,(\phi_{,\mu ;\nu}-g_{\mu\nu}\Box \phi)
  +\frac{8\pi }{c^4\phi}\, T_{\mu\nu}
\end{equation}
for the geometric part, and
\begin{equation}\label{2.2}
  \frac{2\omega}{\phi}\,\Box \phi-\frac{\omega}{\phi^2} \phi_{,\mu}\phi^{,\mu}
  +R=0
\end{equation}
for the scalar field. $\Box$ is the usual d'Alembert operator in
curved space--time and $T_{\mu\nu}$ is the momentum--energy tensor
of matter.

The line element describing a static and isotropic geometry is
expressed as
\begin{equation}\label{2.3}
  ds^2=-e^{2\alpha}dt^2+e^{2\beta}[dr^2+r^2(d\theta^2+\sin^2\theta d\varphi^2)]\,,
\end{equation}
where the functions $\alpha$ and $\beta$ depend on the radial
coordinate $r$. The general solution in the vacuum is given by
\cite{brans}
\begin{eqnarray}
 e^{2\alpha} & = & e^{2\alpha_0}\left[\frac{1-B/r}{1+B/r}\right]^{2/\lambda}\,,\label{2.4} \\
 e^{2\beta} & = & e^{2\beta_0}\left(1+\frac{B}{r}\right)^4
 \left[\frac{1-B/r}{1+B/r}\right]^{2(\lambda -C-1)/\lambda}\,, \label{2.5} \\
 \phi & = & \phi_0\left[\frac{1-B/r}{1+B/r}\right]^{-C/\lambda} \,, \label{2.6}
\end{eqnarray}
where $C$, $\lambda$, $\phi_0$, $\alpha_0$ and $\beta_0$ are
arbitrary constants. $\lambda$ is related to the constant $C$
through the relation
 \begin{equation}\label{lambda}
 \lambda^2=1+C+C^2\left(1+\frac{\omega}{2}\right)\,,
 \end{equation}
and $B$ is related to the mass $M$ of the source. Without to lose
of the generality, we assume $\alpha_0=0$ and $\beta_0=0$. Here
$\omega>3/2$ \cite{brans}.

In the weak field approximation, the components of the metric
tensor, $g_{\mu\nu} \simeq \eta_{\mu\nu}+h_{\mu\nu}$, and the BD
scalar reduce to the form \cite{brans}
\begin{eqnarray}
  g_{00} & \simeq & -1+\frac{2M\phi_0^{-1}}{c^2 r}\frac{4+2\omega}{3+2\omega}\,,
  \label{2.8}\\
  g_{ii} & \sim &  1+\frac{2M\phi_0^{-1}}{c^2 r}\frac{2+2\omega}{3+2\omega}\,,
  \quad i=1,2,3, \label{2.9} \\
  g_{0i} & = & 0\,, \qquad g_{ij}=0\,,\quad i\neq j\,, \label{2.10}\\
  \phi & = & \phi_0 +\frac{2M}{c^2 r}\frac{1}{3+2\omega} \label{2.11} \,.
\end{eqnarray}
In order to have the agreement with observations, Eqs.
(\ref{2.8})-(\ref{2.11}) are obtained by means of an appropriate
choice of the constant entering in Eqs. (\ref{2.4})-(\ref{2.6})
\cite{brans}
\begin{equation}\label{2.7}
  \lambda =\sqrt{\frac{2\omega +3}{2(\omega +2)}}\,, \quad C\cong -\frac{1}{2+\omega}\,,
  \quad \alpha_0=0=\beta_0\,,
\end{equation}
 $$
 \phi_0=\frac{4+2\omega}{G_N(3+2\omega)}\,, \quad B=\frac{M}{2c^2\phi_0}\,
 \sqrt{\frac{2\omega+4}{2\omega+3}}\,.
 $$
In the limit $\omega\to \infty$, the usual weak field
approximation of General Relativity is recovered.

\section{The Cerenkov Effect}
We now investigate the possibility that a charge particle,
propagating in a gravitational field described by BD theory, could
emit radiation through the Cerenkov process. Such a process has
been analyzed some years ago by Gasperini \cite{gasperini}. Here
we follow a different approach proposed in a recent paper by
Gupta, Mohanty and Samal \cite{gupta}: The Cerenkov process occurs
owing to a different coupling of fermions and photons with the
gravitational background. They show that the gravitational field
acts as an effective refractive index, whose expression is given
by \cite{gupta}
\begin{equation}\label{1.1a}
  n^2_T(k_0)=1-\frac{R^i_{\phantom{i}i}}{|\eta^{00}| k_0^2}+
  \frac{k_ik^j}{2{\bf k}^2}\frac{R^i_{\phantom{i}\, j}}{k_0^2|\eta^{00}|}\,,
\end{equation}
for the transverse modes, and
\begin{equation}\label{1.1b}
  n^2_L(k_0)=1-\frac{k_ik^j}{{\bf k}^2|\eta^{00}|}\frac{R^i_{\phantom{i}\, j}}{k_0^2}\,,
\end{equation}
for the longitudinal modes. $R^i_{\phantom{i}i}$ is understood as
the sum over the spatial indices of the Ricci tensor
$R^{\mu}_{\phantom{\mu}\nu}$, i.e.
$R^i_{\phantom{i}i}\equiv\sum_{i=1}^3R^i_{\phantom{i}i}$, $k_0$ is
the frequency of the  emitted photon $\gamma$ ($k^\mu=(k^0, {\bf
k})$, with ${\bf k}=(k^1, k^2, k^3)$), and $\eta^{00}$ is the
00-component of the metric tensor in the inertial frame,
$\eta_{\mu\nu}=(-1, 1,1,1)$.

The scattering process $f(p)\to f(p')+\gamma(k)$, responsible of
the Cerenkov radiation, is analyzed in the local inertial frame
of the incoming fermion $f(p)$ with momentum $p$, while $f(p')$
and $\gamma(k)$ are the outcoming fermion with momentum $p'$ and
the emitted photon with momentum $k$. As one can immediately
realize, the Cerenkov emission is not vanishing in the inertial
frame since the refractive index turns out to be proportional to
the Ricci tensor \cite{gupta}.

The energy radiated via Cerenkov effect by a charged particle
moving in a background gravitational field is given by
\begin{equation}\label{1.2}
  \frac{dE}{dt}=\frac{Q^2\alpha_{em}}{4\pi p_0^2}
  \int_{k_{01}}^{k_{02}}dk_0\left[p_0(p_0-k_0)-\frac{1}{2}k_0^2\right]
  k_0\frac{n_{\gamma}^2(k_0)-1}{n_{\gamma}^2(k_0)}\,,
\end{equation}
where $Q$ is the charge of the fermion emitting the photon,
$\alpha_{em}$ is the electromagnetic coupling constant, and $p_0$
is the energy of the fermion. Here $n_\gamma$ indicates the
refractive index for photons.

The non-vanishing components of the Ricci tensor corresponding to
the metric (\ref{2.3}) are
 \begin{eqnarray}
 R_{00}&=& -\frac{4B^2Cr^4}{(r^2-B^2)^4\lambda^2}\left(\frac{r-B}{r+B}
 \right)^{2(2+C)/\lambda}\,, \label{R0} \\
 R_{11}=R_{rr}&=&-\frac{4B[B^2C\lambda +Cr^2\lambda+Br(2+C-2\lambda^2)]}
 {r(r^2-B^2)^2\lambda^2} \,, \label{R1} \\
 R_{22}=R_{\theta\theta}&=&\frac{2BCr[-2B(1+C)r+B^2\lambda+
                 r^2\lambda]}{(r^2-B^2)^2\lambda^2}\,, \label{R2} \\
 R_{33}=R_{\varphi\varphi}&=&\frac{2BCr[-2B(1+C)r+B^2\lambda+r^2\lambda]\sin^2\theta}
 {(r^2-B^2)^2\lambda^2}\,, \label{R3}
 \end{eqnarray}
whereas the scalar curvature is
\begin{equation}\label{curvature}
  R=\frac{4B^2r^4}{(r^2-B^2)^4\lambda^2}\left(\frac{r-B}{r+B}\right)^{2(1+C)/\lambda}
  C^2\omega\,.
\end{equation}
The summation over the spatial components of Ricci's tensor is
then
\begin{equation}\label{ricci}
  R_i^i=\frac{4B^2r^4}{(r^2-B^2)^4\lambda^2}\left(\frac{r-B}{r+B}\right)^{2(1+C)/\lambda}
  C(-1+C\omega)\,.
\end{equation}
The crucial point in order for the Cerenkov radiation to be
kinematically allowed is that the refractive index (Eqs.
(\ref{1.1a}) and/or (\ref{1.1b})) is greater than 1. To show that
this occurs in BD theory, we shall discuss some particular cases.

\begin{description}

\item[-] If the momentum of photons is directed as ${\bf k}=(k, 0, 0)$,
and $r\gg B$, then the
  last term in (\ref{1.1a}) and (\ref{1.1b}) reads
\begin{equation}\label{last}
  \frac{k_ik^j}{{\bf k}^2}\frac{R^i_j}{k_0^2}\sim
  \frac{R_{11}}{k_0^2}\sim -\frac{4BC}{r^3\lambda k_0^2}\sim
  \frac{4M}{r^3c^2\phi_0 k_0^2}\frac{1}{2\omega +3}\,,
\end{equation}
where Eqs. (\ref{2.7}) have been used. Similarly, one gets
\begin{equation}\label{ricci-weak}
  R_i^i\sim
  \frac{8M^2}{r^4c^4\phi_0^2}\frac{\omega+1}{(2\omega+3)^2}\,.
\end{equation}
Therefore, Eq. (\ref{1.1a}) assumes the form
\begin{equation}\label{nT}
  n_T^2\sim 1+\frac{4M}{r^3c^2\phi_0k_0^2}\frac{1}{2\omega +3}-
  \frac{8M^2}{r^4c^4\phi_0^2k_0^2}\frac{\omega+1}{(2\omega+3)^2}\,,
\end{equation}
which is greater than 1 being last term of the order ${\cal
O}(r^{-4})$. The longitudinal refractive index turns out to be
lesser than 1, as follows from (\ref{1.1b}).

\item[-] As before observed, $B$ is a constant related to the source
mass, and $C$ is an arbitrary constant entering in the BD field
equations describing the gravitational field (Eqs.
(\ref{2.4})-(\ref{2.6})). The weak field solutions, Eqs.
(\ref{2.8})-(\ref{2.11}), are permissible {\it only} if the
gravitational source is a suitable mass distribution, as for
example the Sun, which generates everywhere a small field, inside
and outside to it (see Eq. (\ref{nT})). This does not hold for a
point mass source or, in the more general case, for high density
of matter \cite{brans}. Let us consider the gravitational field
generated by a source placed at the center of Galaxies. In such a
case, the field weak approximation cannot be trivially used, at
least in the BD conjecture, because of the strong field regime
inside the gravitational source. This situation avoids of using
the constraints (\ref{2.7}). Thus, as follows from (\ref{1.1a}),
(\ref{1.1b}) and (\ref{ricci}), $n_T^2>1$ provided $R_i^i$ is
negative and the last term in (\ref{1.1a}) and (\ref{1.1b}) is
positive for $\omega>0$, $C>0$ and for $r$ sufficiently large with
respect to $B$. The condition $R_i^i<0$ implies the constraint
$\omega C<1$. As expected, the effect disappears for $C=0$, i.e.
when the static solution (\ref{2.4})-(\ref{2.5}) reduces to the
solution of Einstein's theory.

\item[-] One can also investigate the case $\lambda \gg 1$
($\omega\gg 1$) and $r\gg B$. Eqs. (\ref{R1})-(\ref{R3}) and
(\ref{ricci}) becomes
 \[
 R_{11}\sim -\frac{B}{r^3\sqrt{\omega}}\, \quad R_{22}\sim
 \frac{B}{r\sqrt{\omega}}\,, \quad R_{33}\sim
 \frac{B\sin\theta}{r\sqrt{\omega}}\,, \quad R_i^i\sim
 \frac{B^2}{r^4}\,.
 \]
Notice these expressions do not depend on $C$. For photons
propagating with momentum ${\bf k}=(k, 0, 0)$ the refractive
indices are given by
 \[
 n_T^2\sim 1+\frac{B}{r^3k_0^2\sqrt{\omega}}-
 \frac{B^2}{r^4k_0^2}>1\,,
 \quad
  n_L^2\sim 1-\frac{B}{r^3k_0^2\sqrt{\omega}}<1\,.
 \]
\end{description}
The above results clearly show that the BD theory of gravity is a
suitable framework for the occurrence of the Cerenkov effect,
showing up a departure from General Relativity. Nevertheless, it
must be noted that large bounds on $\omega$ yield a refractive
index of the order $n_\gamma^2\sim {\cal O}(1)$. Thus, from an
experimental setting, the Cerenkov process is very difficult to
be detected.

\section{Discussion and Conclusions}
In this paper we have analyzed, in the framework of BD theory of
gravity, the Cerenkov emission by a charged particle. Since the
Cerenkov radiation occurs for particles with very high energy, the
suitable context for producing it could be represented by
astrophysical and cosmological scenarios. For example, one can
envisage the studying of astrophysical objects present at the
center of Galaxies. Let us recall that it is widely believed that
the central mass of Galaxies has to be a black hole described by
General Relativity. But in our own galaxy, for instance, if Sgr
A$^*$ (the super-massive compact object name) is a black hole, its
luminosity should be three order of magnitude bigger than that
which is observed. This discrepancy is called {\it the blackness
problem}. In addition, observational data come from regions at a
radius larger than $4 \times 10^{4}$ Schwarzschild radius of a
black hole of mass $2.6 \times 10^{6}$ solar masses -which is the
inferred mass of the central object-, and so, proofs of the
existence of a super-massive black hole in the center of the
galaxy are not {\it conclusive} (alternative models, in fact, have
been proposed in \cite{diego,violler}). It is then of current
interest to edit the analysis of a super-massive star placed at
the center of galaxies whose gravitational field could satisfy the
requirement of Mach's principle and look for Cerenkov's radiation
emitted by accelerated charged particles, which is an allowed
process in the BD theory of gravity, but vanishes in the framework
of General Relativity. As a consequence, the detection of the
Cerenkov radiation might give a strong signature in favour of BD
theory.

It would be interesting to extend this analysis to scalar
tensor-theories theories of gravity (as well as higher order
theories of gravity), in which the strength of the coupling
between the scalar field and gravity is determined by the function
$\omega (\phi )$. In the more general case, a self--interaction
potential $V(\phi)$, which plays a non trivial role on the
dynamics of the field, can be also introduced. The dependence of
the parameter $\omega$ on $\phi$ could have the property that, at
the present epoch, the value of the scalar field $\phi_0$ is such
that $\omega$ is very large, leading to theories almost identical
to General Relativity today, but for past or future values of
$\phi$, $\omega$ could take values that would lead to significant
differences from General Relativity, hence to an enhancement of
Cerenkov effect. Such extension is currently under investigation.

\acknowledgments Research supported by MURST PRIN 99. The author
want express his thankfulness to referees whose comments and
suggestions have improved the paper.

\end{document}